\documentclass{PoS}

\title{STAR's latest results on quarkonium production}

\ShortTitle{STAR's latest results on quarkonium production}

\author{\speaker{Barbara Trzeciak (for the STAR Collaboration)}%
        \\Faculty of Nuclear Sciences and Physical Engineering \\
		Czech Technical University in Prague \\
		Brehova 7, 115 19 Prague 1, Czech Republic \\
        E-mail: \email{trzecbar (at) fjfi.cvut.cz}}

\abstract{
In this proceedings, we report the latest results of J/$\psi$ and $\Upsilon$ production from the STAR experiment at RHIC, in different colliding systems and colliding energies. J/$\psi$ nuclear modification factors ($R_{AA}$) in Au+Au collisions at $\sqrt{s_{NN}} =$ 200, 62.4 and 39 GeV and in U+U collisions at $\sqrt{s_{NN}} =$ 193 GeV, $\Upsilon$ $R_{AA}$ in $d+$Au and Au+Au collisions at $\sqrt{s_{NN}} =$ 200 and in U+U collisions at $\sqrt{s_{NN}} =$ 193 GeV are shown and compared to different theoretical models. We also present prospects of quarkonium measurements at STAR.}

\FullConference{53rd International Winter Meeting on Nuclear Physics\\
                 26-30 January 2015\\
                 Bormio, Italy}

\begin{document}
 
\section{Introduction}

It was proposed that in high energy heavy-ion collisions quarkonium states can be used as probes of the Quark-Gluon Plasma (QGP) formation. Due to the Debye-like color screening of the quark-antiquark potential in the hot and dense medium, quarkonia are expected to dissociate and this ''melting'' can be a signature of the presence of a QGP~\cite{Matsui:1986dk}. Moreover, different quarkonium states have different binding energies and thus sizes and so they are expected to melt at different temperatures in the QGP. Therefore, this sequential quarkonium suppression pattern can be used as a QGP thermometer~\cite{Mocsy:2008eg}. Studies of production of quarkonium states in heavy-ion collisions can provide insight into the thermodynamic properties of the hot and dense medium~\cite{Mocsy:2007jz}.

But there are other mechanisms that can alter quarkonium yields in heavy-ion collisions relative to $p+p$ collisions.
In the hot medium the yields can be enhanced due to statistical recombination of heavy quark-antiquark pairs. Also, effects related to the ''normal'' nuclear matter, so called cold nuclear matter (CNM) effects - such as shadowing, Cronin effect, initial-state parton energy loss or final state nuclear absorption - can affect qarkonium production.
Due to the interplay of all mentioned mechanisms, it is difficult to isolate the color screening effect from other effects. At RHIC energies, $\Upsilon$ are considered as cleaner probes of the QGP, compared to J/$\psi$, because of the negligible statistical recombination and co-mover absorption~\cite{Rapp:2008tf,Adamczyk:2013poh}. J/$\psi$ with high-$p_{T}$, $>$ 5 GeV/$c$, are also expected to be almost not affected by the recombination and CNM effects~\cite{Zhao:2010nk}. STAR measurement of the J/$\psi$ elliptic flow ($v_{2}$)~\cite{Adamczyk:2012pw} confirmed negligible recombination, and data disfavor the scenario that J/$\psi$ are dominantly produced by statistical recombination from thermalized $c\bar{c}$ pairs, at $p_{T} >$ 2 GeV/$c$.

Interpretation of experimental results is also complicated because the significant fraction of J/$\psi$ and $\Upsilon$ originate from feed-down sources. Inclusive J/$\psi$ production is a combination of prompt and non-prompt J/$\psi$. The prompt J/$\psi$ production consists of the direct one ($\sim$60\%) and of a feed-down from excited states $\psi (2S)$ and $\chi _{C}$, while non-prompt J/$\psi$ originate from B-hadron decays. The feed-down contribution to the production of the ground state $\Upsilon(1S)$ from excited state is measured to be $\sim$ 30-50\%~\cite{Affolder:1999wm}. 

Systematic measurements of quarkonium production as a function of centrality and transverse momentum, for different colliding systems and collision energies may help to understand the quarkonium production mechanisms in heavy-ion collisions as well as properties of the created medium.
We present here STAR results on J/$\psi$ production in Au+Au collisions at $\sqrt{s_{NN}} =$ 200, 62.4 and 39 GeV and in U+U collisions at $\sqrt{s_{NN}} =$ 193 GeV. 
Also, $\Upsilon (1S+2S+3S)$ and $\Upsilon (1S)$ measurements in $p+p$, $d$+Au and Au+Au collisions at $\sqrt{s_{NN}} =$ 200 GeV and in U+U collisions at $\sqrt{s_{NN}} =$ 193 GeV are presented. Finally, we give prospects of the quakonium measurements in STAR with the recent upgrades.

\section{Quarkonium measurements with the STAR experiment}
\label{sec:measurements}

In STAR, different quarkonium states - J/$\psi$, $\psi(2S)$ and $\Upsilon$ states - have been analysed via their di-electron decay channels. The STAR detector~\cite{Ackermann:2002ad} is a multi-purpose detector that has a large acceptance at mid-rapidity, $\vert \eta \vert <$ 1, with a full azimuthal coverage. 
The Time Projection Chamber (TPC)~\cite{Anderson:2003ur} is the main tracking system and is used to identify particles through the ionization energy loss ($dE/dx$) measurement.
Furthermore, electrons can be selected using the Time Of Flight (TOF) detector~\cite{Llope:2012zz} that greatly enhances the electron identification capability at low momentum where the $dE/dx$ bands for electrons and hadrons overlap. 
High-$p_{T}$ electron identification can be improved by the Barrel Electromagnetic Calorimeter (BEMC)~\cite{Beddo:2002zx} which measures energy deposited in the detector. The BEMC is also used to trigger on high-$p_{T}$ electrons (HT trigger). Minimum bias (MB) data are triggered by the Vertex Position Detectors (VPD)~\cite{Llope:2003ti}.

\subsection{Results on J/$\psi$ production}
\label{sec:JpsiMeasurements}

In order to evaluate effects of the medium created in heavy-ion collisions on J/$\psi$ production, the nuclear modification factors ($R_{AA}$) in Au+Au and U+U collisions were calculated. $R_{AA}$ is defined as a ratio of the particle yield in $A+A$ collisions to that in $p+p$ collisions, scaled by the number of binary collisions. The measurements were done as a function of transverse momentum, and for different centrality bins, which are represented as the number of participant nucleons ($N_{part}$) in a collision.

\begin{figure}[ht]
		\centering
		\includegraphics[width=0.49\textwidth]{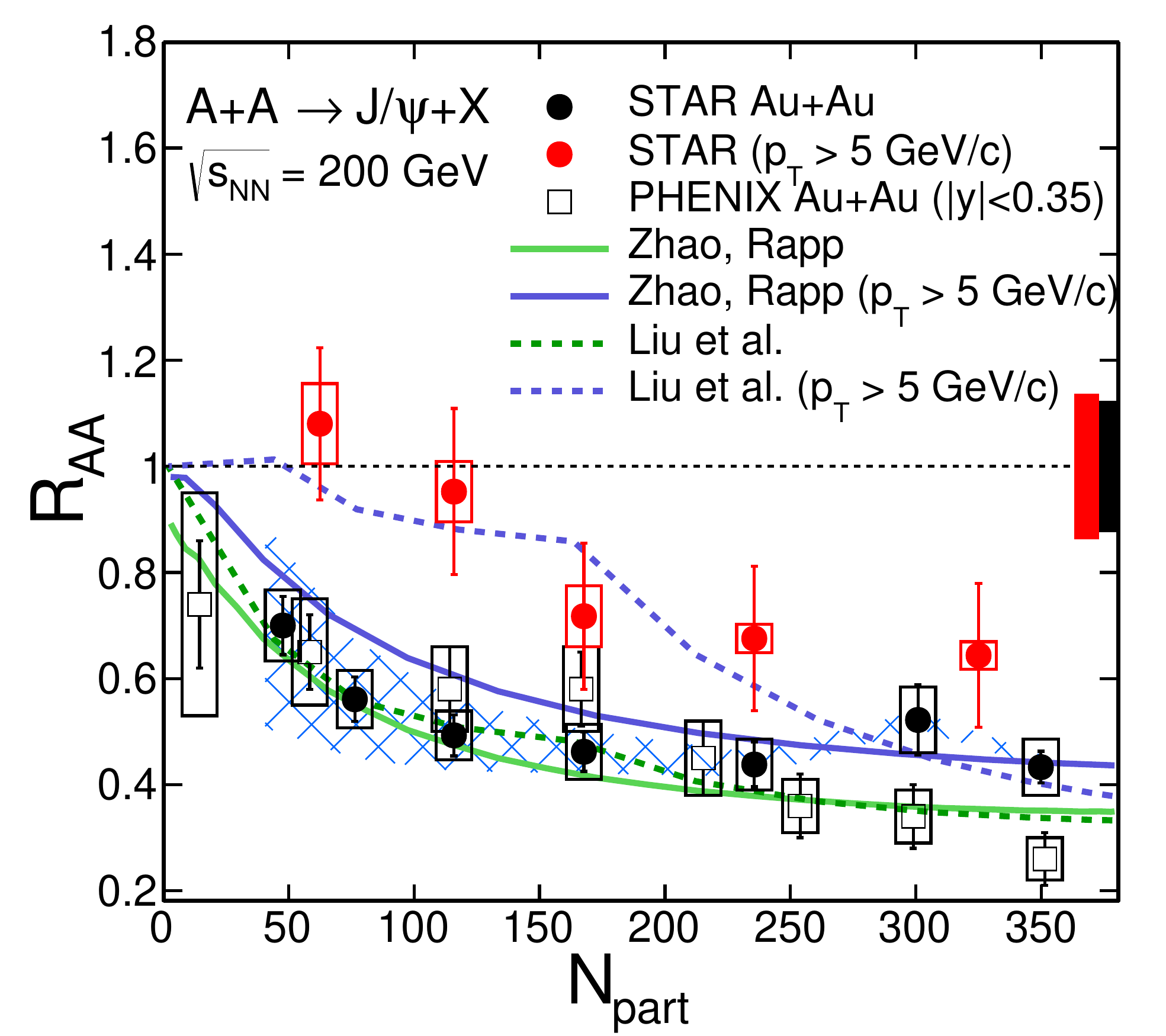}
		\includegraphics[width=0.49\linewidth]{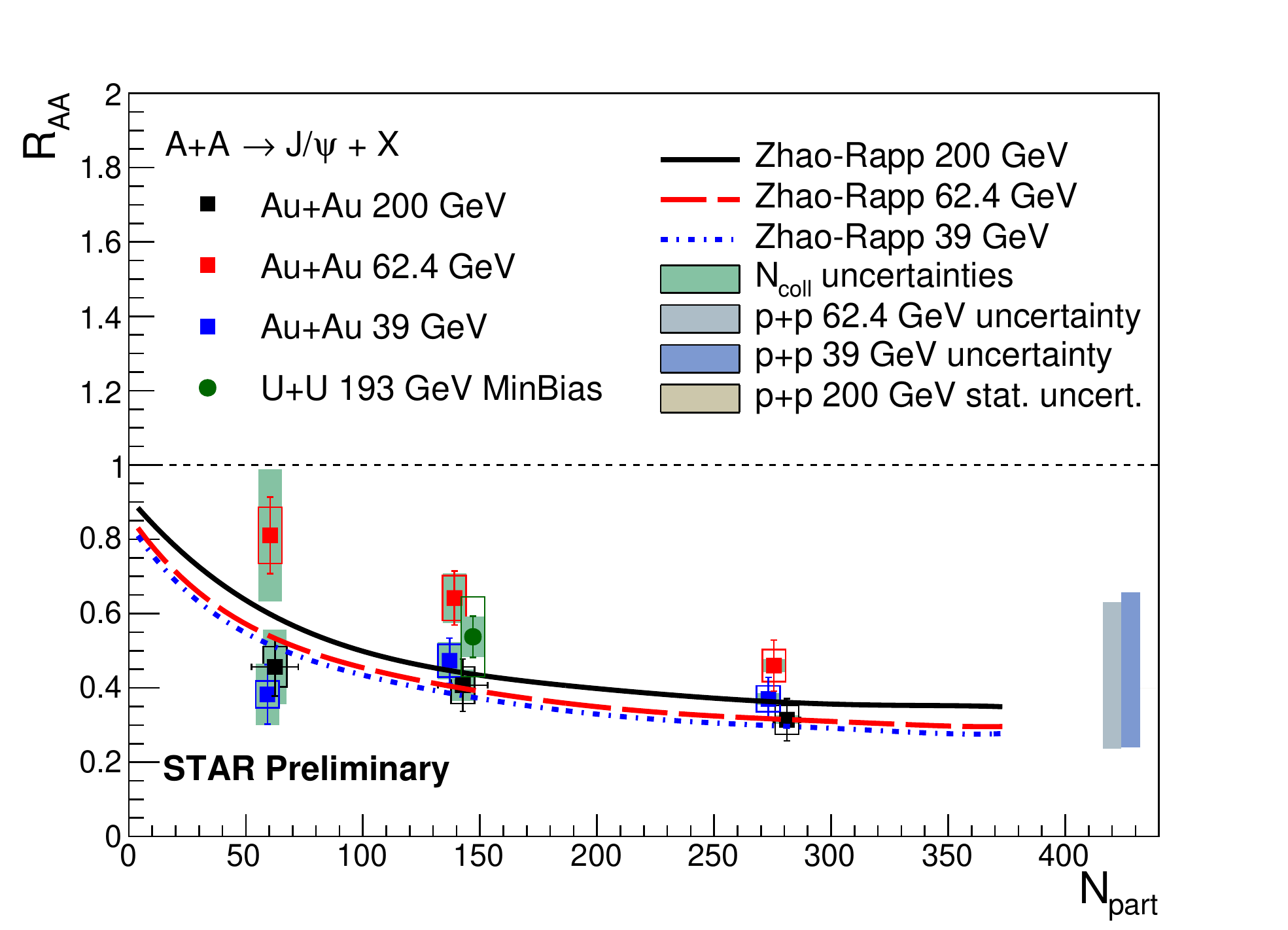}
		\caption{Left: J/$\psi$ $R_{AA}$ as a function of $N_{part}$ in Au+Au collisions at $\sqrt{s_{NN}} =$ 200 GeV at mid-rapidity (\cite{Adamczyk:2012ey,Adamczyk:2013tvk}) with two model predictions (\cite{Zhao:2010nk,Liu:2009nb}). The low-$p_{T}$ ($<$ 5 GeV/$c$) result is shown as black full circles and the high-$p_{T}$ ($>$ 5 GeV/$c$) measurement as red full circles. Right: J/$\psi$ $R_{AA}$ as a function of $N_{part}$ in Au+Au collisions at $\sqrt{s_{NN}} =$ 200 (black), 62.4 (red) and 39 (blue) GeV at mid-rapidity with model predictions. As the green circle the minimum bias U+U measurement at $\sqrt{s_{NN}} =$ 193 GeV is also presented. }
		\label{fig:Jpsi_raa}
\end{figure}

The left panel of Fig.~\ref{fig:Jpsi_raa} shows J/$\psi$ $R_{AA}$ in Au+Au collisions at $\sqrt{s_{NN}} =$ 200~GeV as a function of $N_{part}$, for low- ($<$ 5 GeV/$c$)~\cite{Adamczyk:2013tvk} and high-$p_{T}$ ($>$ 5 GeV/$c$)~\cite{Adamczyk:2012ey} ranges. J/$\psi$ suppression increases with the collision centrality, and the suppression level is systematically lower for high-$p_{T}$ J/$\psi$ compared to low-$p_{T}$ ones. 
Suppression of high-$p_{T}$ J/$\psi$ observed in central collisions (0-30\%) points to the color screening effect and formation of the QGP, since high-$p_{T}$ J/$\psi$ are expected to be not affected by the recombination or CNM effects.

STAR results are compared with two model predictions, Zhao and Rapp~\cite{Zhao:2010nk} and Liu {\it{et al.}}~\cite{Liu:2009nb}. Both take into account direct J/$\psi$ production with the color screening effect and production via recombination of $c$ and $\bar{c}$ quarks. The Zhao and Rapp model also includes the J/$\psi$ formation time effect and the B-hadron feed-down contribution. At low $p_{T}$ both predictions (green lines) are in agreement with the data, while the high-$p_{T}$ result is well described by the Liu {\it{et al.}} model and the model of Zhao and Rapp underpredicts the measured $R_{AA}$. The high-$p_{T}$ predictions are represented by blue lines in Fig.\ref{fig:Jpsi_raa}.

One can study the interplay between recombination, CNM effects and direct J/$\psi$ production (that can be influenced by the color screening effect) by changing energies of colliding ions.
The right panel of Fig.~\ref{fig:Jpsi_raa} shows low-$p_{T}$ J/$\psi$ $R_{AA}$ in Au+Au collisions for different colliding energies, $\sqrt{s_{NN}} =$ 200 (black), 62.4 (red) and 39 (blue) GeV. For all these three energies, the suppression is similar within the uncertainties, and results are well described by the model of Zhao and Rapp~\cite{Zhao:2010nk}.
It should be however noted here that due to lack of precise $p+p$ measurements at 62.4 and 39 GeV Color Evaporation Model calculations~\cite{Nelson:2012bc} are used as baselines. This introduces big uncertainties that are shown as boxes in the right panel of Fig.\ref{fig:Jpsi_raa}, separately for each energy.

STAR has also performed an analysis of J/$\psi$ $R_{AA}$ in U+U collisions at $\sqrt{s_{NN}} =$ 193 GeV. The centrality integrated result is shown in the right panel of Fig.\ref{fig:Jpsi_raa}, as a full circle. Energy density reached in U+U collisions can be up to 20\% higher than in Au+Au collisions, in the same centrality bin~\cite{Kikola:2011zz}. In this case, the observed suppression is consistent with that in Au+Au collisions at 200 GeV.

\begin{figure}[ht]
		\centering
		\includegraphics[width=0.49\textwidth]{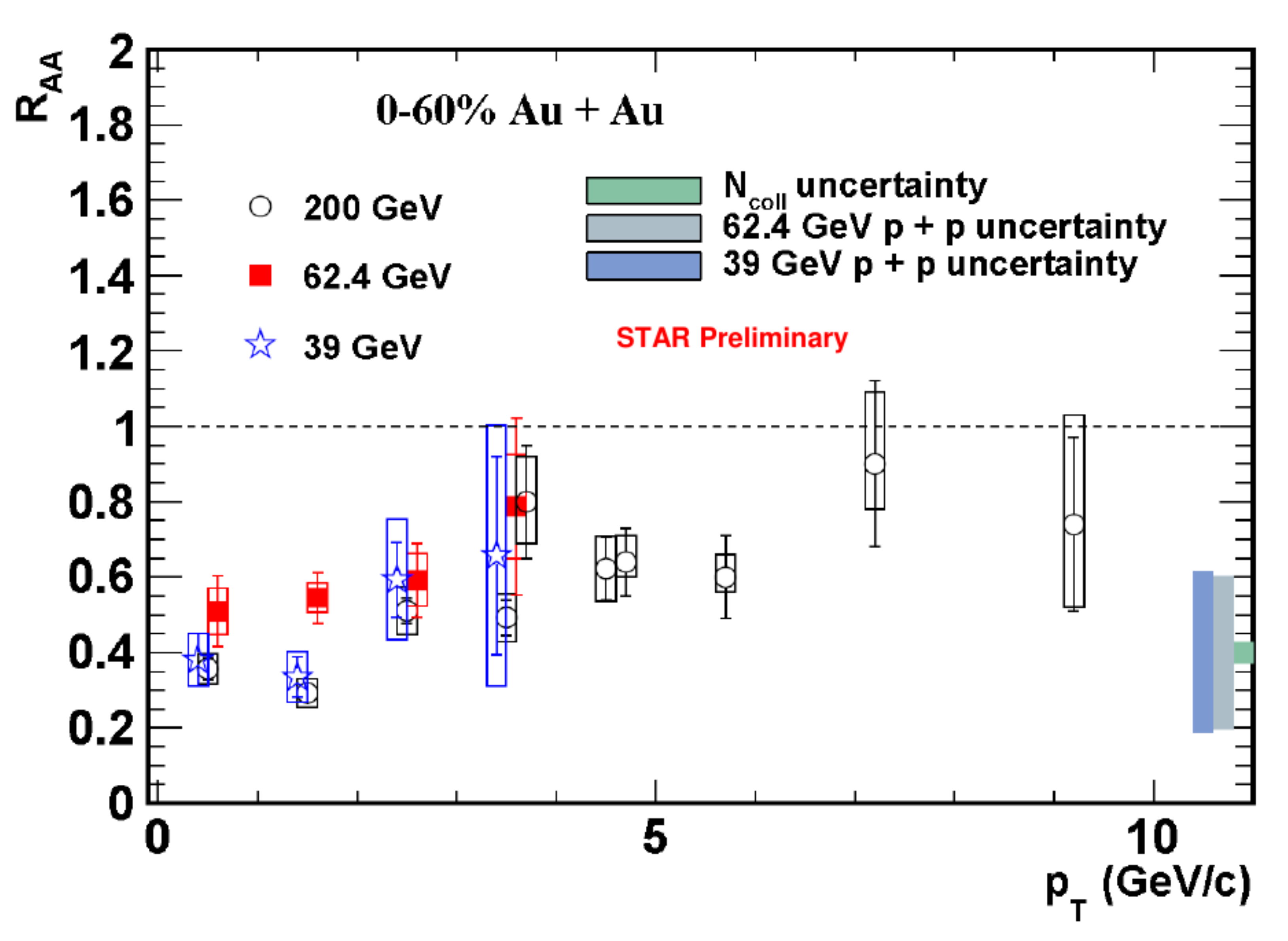}
		\includegraphics[width=0.49\linewidth]{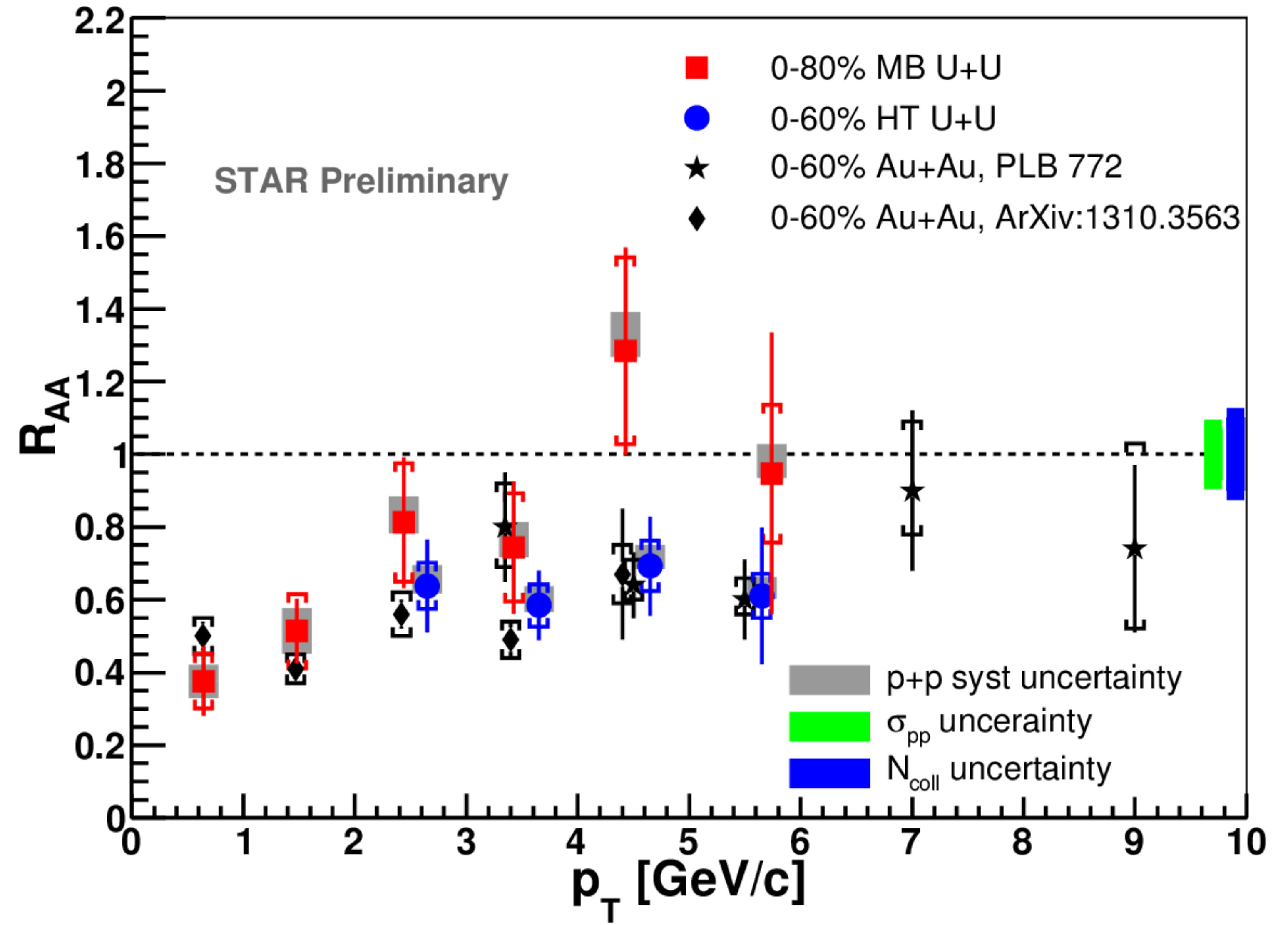}
		\caption{Left: J/$\psi$ $R_{AA}$ as a function of $p_{T}$ in Au+Au collisions at $\sqrt{s_{NN}} =$ 200, 62.4 and 39 GeV at mid-rapidity for min-bias collisions. Right: J/$\psi$ $R_{AA}$ as a function of $p_{T}$ in min-bias Au+Au collisions at $\sqrt{s_{NN}} =$ 200 GeV and U+U at $\sqrt{s_{NN}} =$ 193 GeV. }
		\label{fig:Jpsi_raa_pt}
\end{figure}

As seen in Fig.~\ref{fig:Jpsi_raa_pt}, where nuclear modification factors are presented as a function of transverse momentum, there is almost no differences in $R_{AA}$ vs $p_{T}$ between results from Au+Au collisions at different energies (the left panel), and between Au+Au collisions at $\sqrt{s_{NN}} =$ 200 GeV and U+U collisions at $\sqrt{s_{NN}} =$ 193 GeV (the right panel), within the uncertainties. 

\subsection{Results on $\Upsilon$ production}
\label{sec:UpsilonMeasurements}

Measurements of $\Upsilon$ production in heavy ion collisions help to further investigate the color screening effect. Before presenting results from Au+Au and U+U collisions we show $\Upsilon$ measurements in $p+p$ and $d+$Au collisions, as the baseline and knowledge of the CNM effects are needed. Modification of $\Upsilon$ production in $d+$Au collision, where hot and dense medium is not expected to be created, provides information about CNM effects that can still affect $\Upsilon$, for example shadowing of the parton distribution functions in nucleus or parton energy loss.

\begin{figure}[ht]
		\centering
		\includegraphics[width=0.49\textwidth]{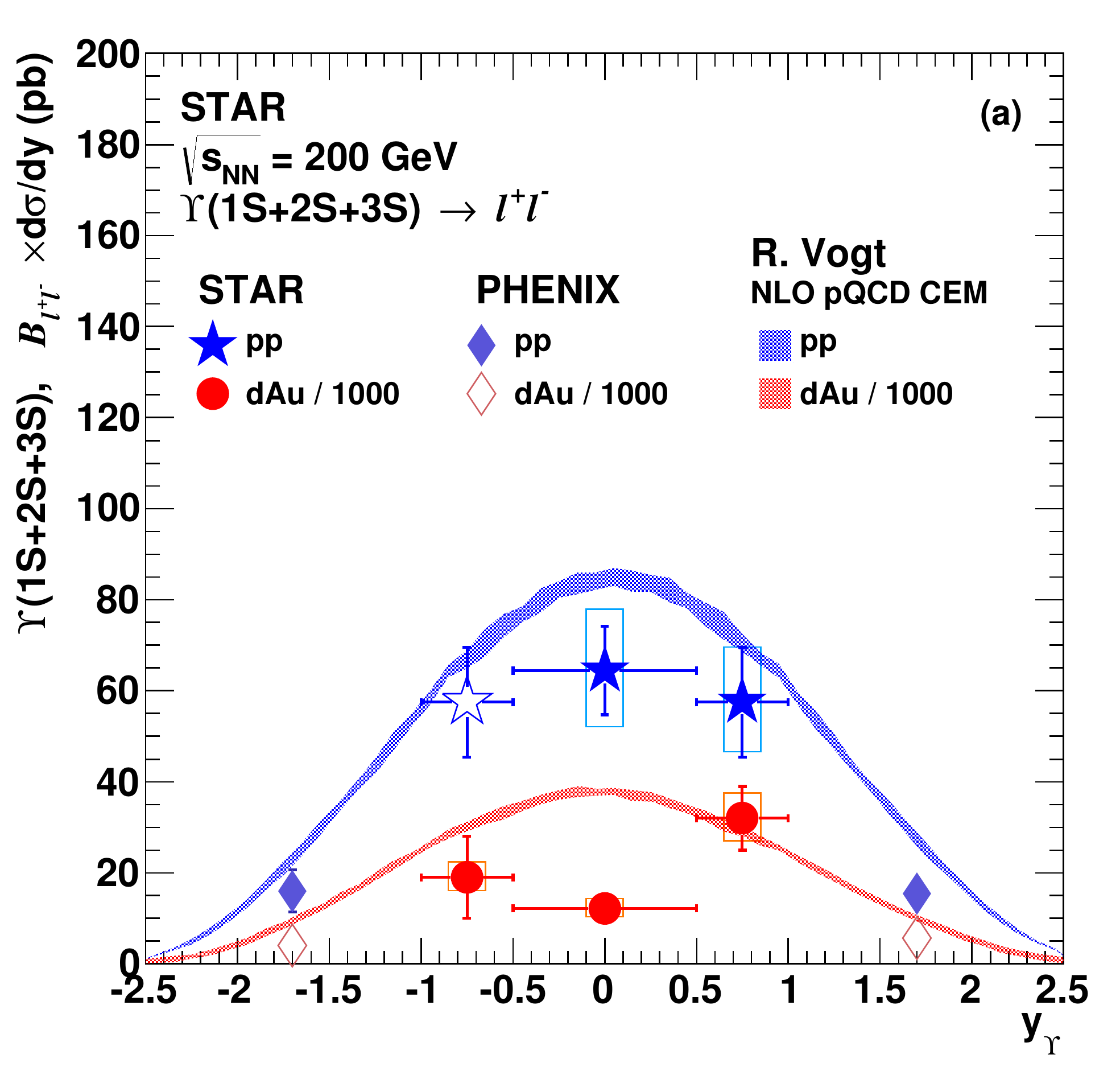}
		\includegraphics[width=0.49\linewidth]{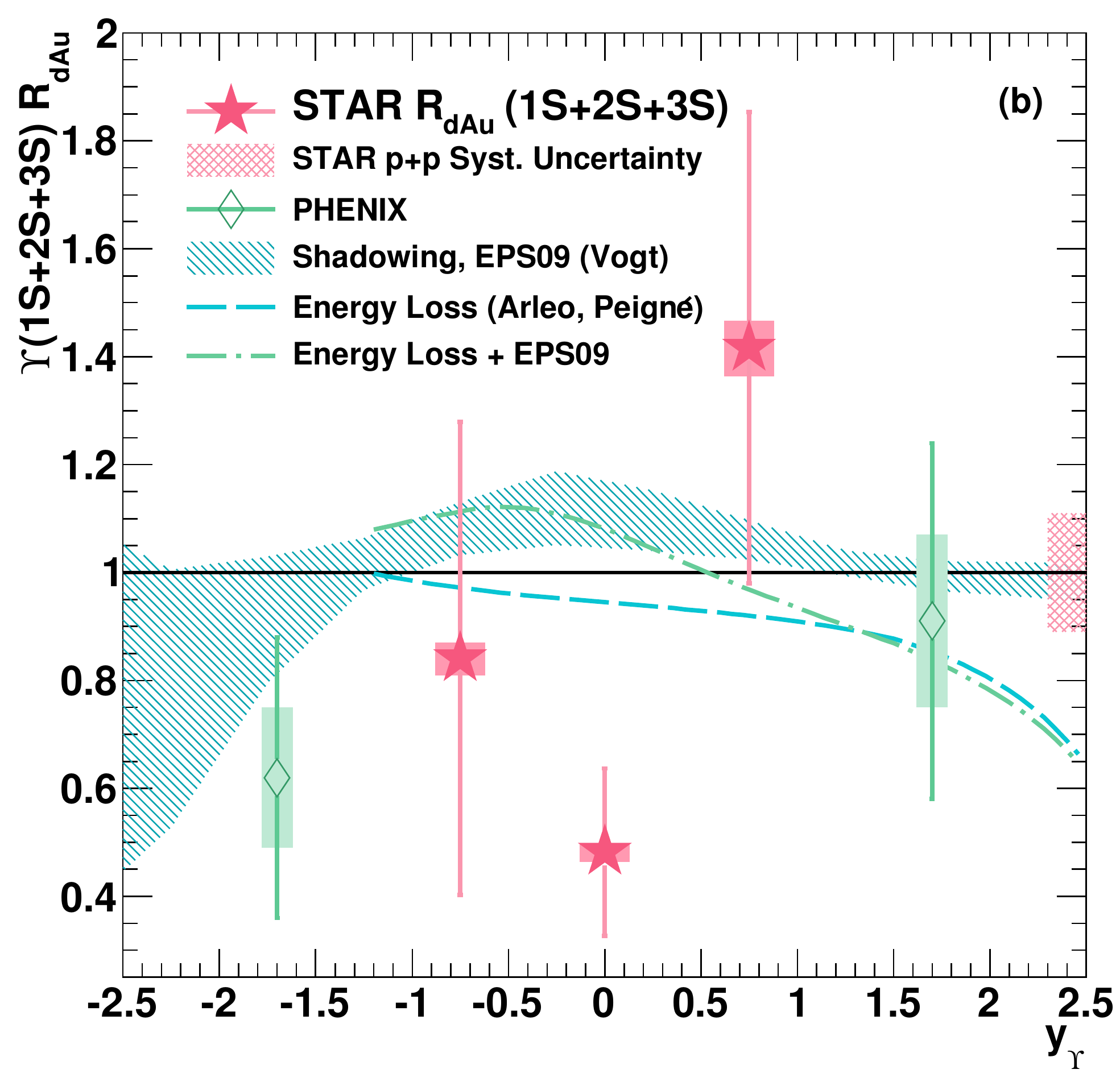}
		\caption{Left: $\Upsilon$ $p+p$ (blue stars) and $d+$Au (red circles) cross-sections (\cite{Adamczyk:2013poh}) as a function of rapidity, compared to CEM NLO pQCD calculations (\cite{Frawley:2008kk}). Right: $R_{dAu}$ as a function of rapidity for STAR (\cite{Adamczyk:2013poh}), red stars, and PHENIX (\cite{Adare:2012bv}), green diamond, results, compared to different model predictions (\cite{Arleo:2012rs}).}
		\label{fig:upsilonppdAu}
\end{figure}

STAR has measured $\Upsilon$ production in $p+p$, $d$+Au and Au+Au collisions at $\sqrt{s_{NN}} =$ 200 GeV~\cite{Adamczyk:2013poh} and in U+U collisions at $\sqrt{s_{NN}} =$ 193 GeV. Figure~\ref{fig:upsilonppdAu} shows $p+p$ and $d+$Au cross-sections as a function of rapidity, represented by stars and full circles, respectively. Measurements of the $p+p$ cross-section, and backward and forward $d+$Au cross-sections are in agreement with NLO pQCD Color Evaporation Model calculations~\cite{Frawley:2008kk}. In the case of $d+$Au, the NLO pQCD CEM calculation takes into account the shadowing effect. There is however a discrepancy at mid-rapidity ($y \sim 0$) region for the $d+$Au measurement which suggests some other effects beside the nuclear PDF modification influencing $\Upsilon$ production at mid-rapidity in $d+$Au collisions. 

The right panel of Fig. ~\ref{fig:upsilonppdAu} presents the nuclear modification factor for $d+$Au collisions, $R_{dAu}$, as a function of rapidity. In order to further evaluate the CNM effects on $\Upsilon$ production, the data are compared to CEM calculations with shadowing based on the EPS09 nPDF parametrization, presented as the shaded area, Arleo {\it{et al.}} model~\cite{Arleo:2012rs} where suppression of $\Upsilon$ is due to initial-state parton energy loss, presented as the dashed line, and the model combining both shadowing and energy loss, presented as the dashed-dotted line. The strong $\Upsilon$ suppression at $y \sim 0$ observed by STAR cannot be explained by the available predictions of CNM effects.
Understanding of these effects is important for the interpretation of results from heavy-ion collisions.

\begin{figure}[ht]
		\centering
		\includegraphics[width=0.49\textwidth]{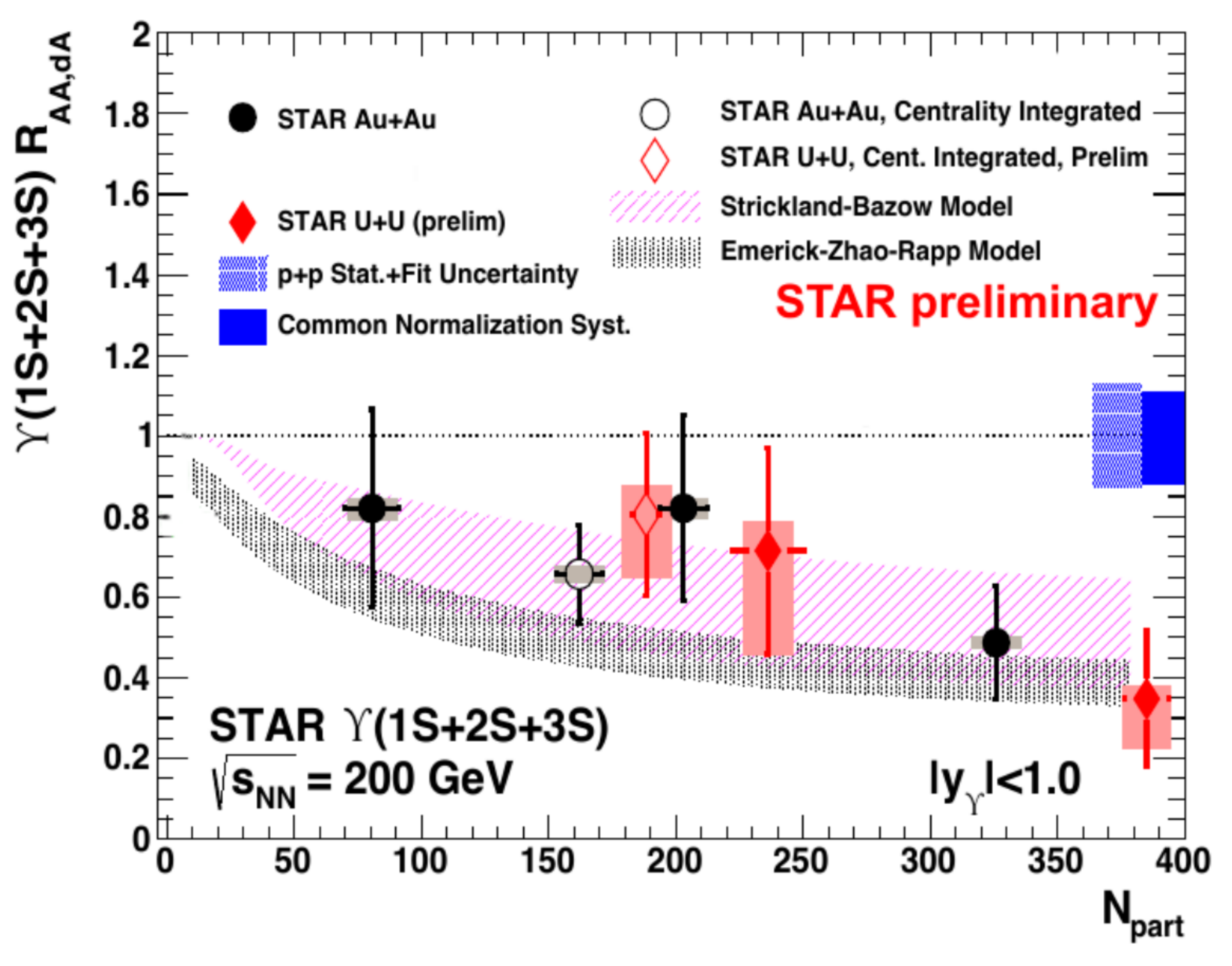}
		\includegraphics[width=0.49\linewidth]{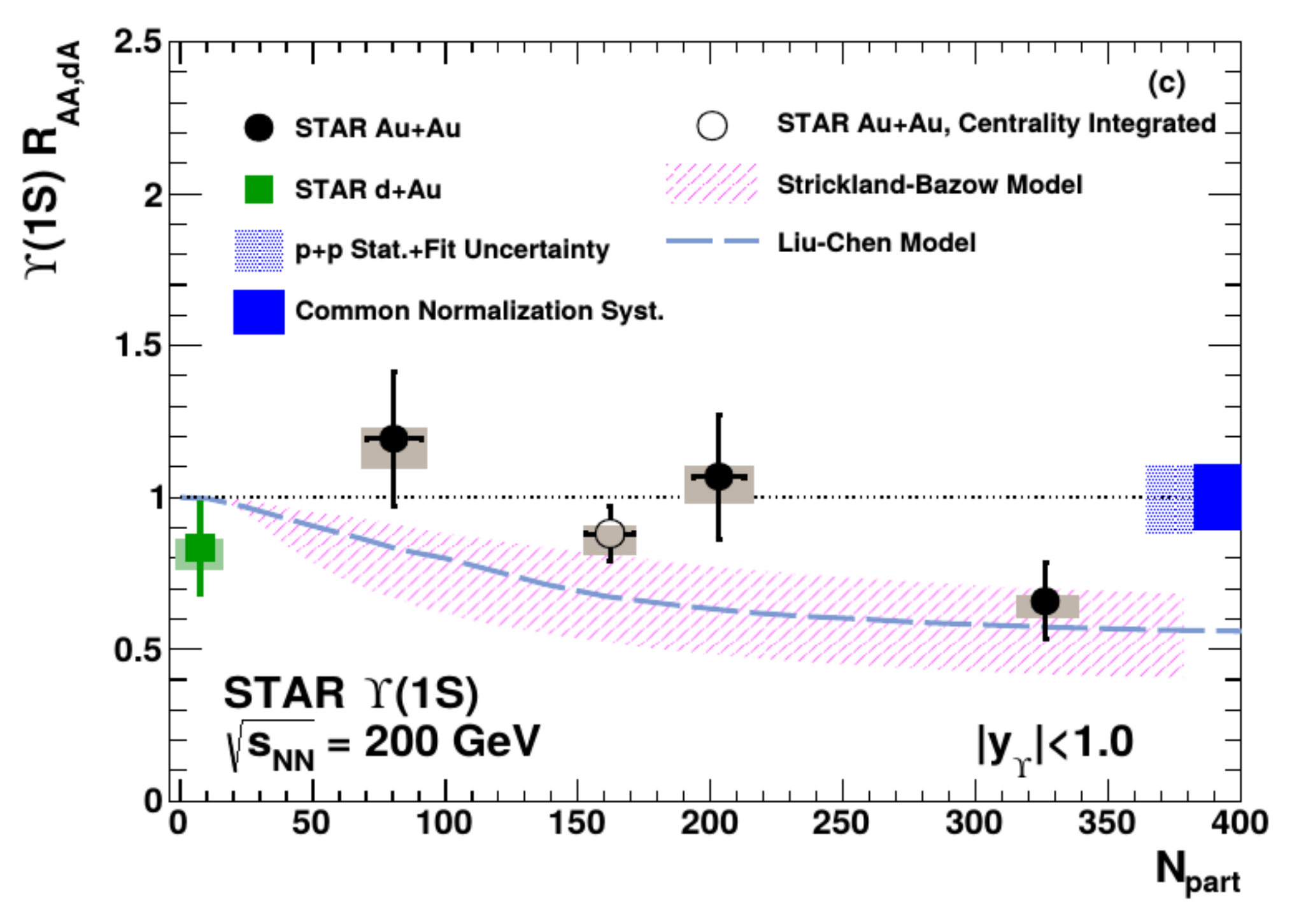}
		\caption{Left: $R_{AA}$ as a function of $N_{part}$ for $\Upsilon(1S+2S+3S)$ at $\vert y \vert < 1$, Au+Au (black circles), (\cite{Adamczyk:2013poh}), and U+U (red diamonds) collisions, compared to two model predictions (shaded areas), \cite{Strickland:2011aa, Emerick:2011xu}. Right: $R_{dA}$ and $R_{AA}$ vs $N_{part}$ for $\Upsilon(1S)$ ground state at $\vert y \vert < 1$ for $d+$Au (green rectangle) and Au+Au (black circles) collisions, (\cite{Adamczyk:2013poh}), compared to different model predictions (shaded area and dashed line), \cite{Strickland:2011aa,Liu:2010ej}.}
		\label{fig:Upsilon_raa}
\end{figure}

The nuclear modification factors for $\Upsilon(1S+2S+3S)$ in Au+Au collisions at $\sqrt{s_{NN}} =$ 200 GeV and U+U collisions at $| y | <$ 1 and $\sqrt{s_{NN}} =$ 193 GeV are shown on the left panel of Fig.~\ref{fig:Upsilon_raa}, as a function of $N_{part}$. The Au+Au and U+U points for different centralities are shown as black circles and red diamonds, respectively.
In the most central collisions the strong suppression is observed: $R_{AuAu} = $0.49$\pm$0.13(Au+Au stat.)$\pm$0.07($p+p$ stat.)$\pm0.02$(Au+Au syst.)$\pm 0.06$($p+p$ syst.) and $R_{UU} = $0.35$\pm$0.17(stat.)$^{\pm0.03}_{-0.13}$(syst.). It is also seen that Au+Au and U+U measurements follow the same trend.
The data are in agreement with two model predictions, of Strickland Emerick {\it{et al.}}~\cite{Emerick:2011xu} and Bozow~\cite{Strickland:2011aa}. Both include hot-nuclear-matter effects, and Emerick {\it{et al.}} calculations include in addition the CNM effects.

With the available $\Upsilon$ statistics it was also possible to separate $\Upsilon(1S)$ from the excited states, $\Upsilon(2S+3S)$. The right panel of Fig.~\ref{fig:Upsilon_raa} presents nuclear modification factors for $\Upsilon(1S)$ in $d+$Au and Au+Au collisions, as a function of centrality. For $d+$Au and Au+Au peripheral collisions, the nuclear modification factor is consistent with unity, while the most central 10\% Au+Au data show suppression. 
The most central Au+Au point is consistent with the prediction of the Liu {\it{et al.}} model~\cite{Liu:2010ej} for inclusive $\Upsilon(1S)$ $R_{AA}$. $\Upsilon(1S)$ suppression within this model is mostly due to the dissociation of the excited states.
For the $\Upsilon(2S+3S)$ the 95\%-confidence upper limit for $R_{AA}$ in the centrality range of 0-60\% was obtained, $R_{AA}(2S+3S) <$ 0.32. 

\begin{figure}[ht]
		\centering
		\includegraphics[width=0.49\linewidth]{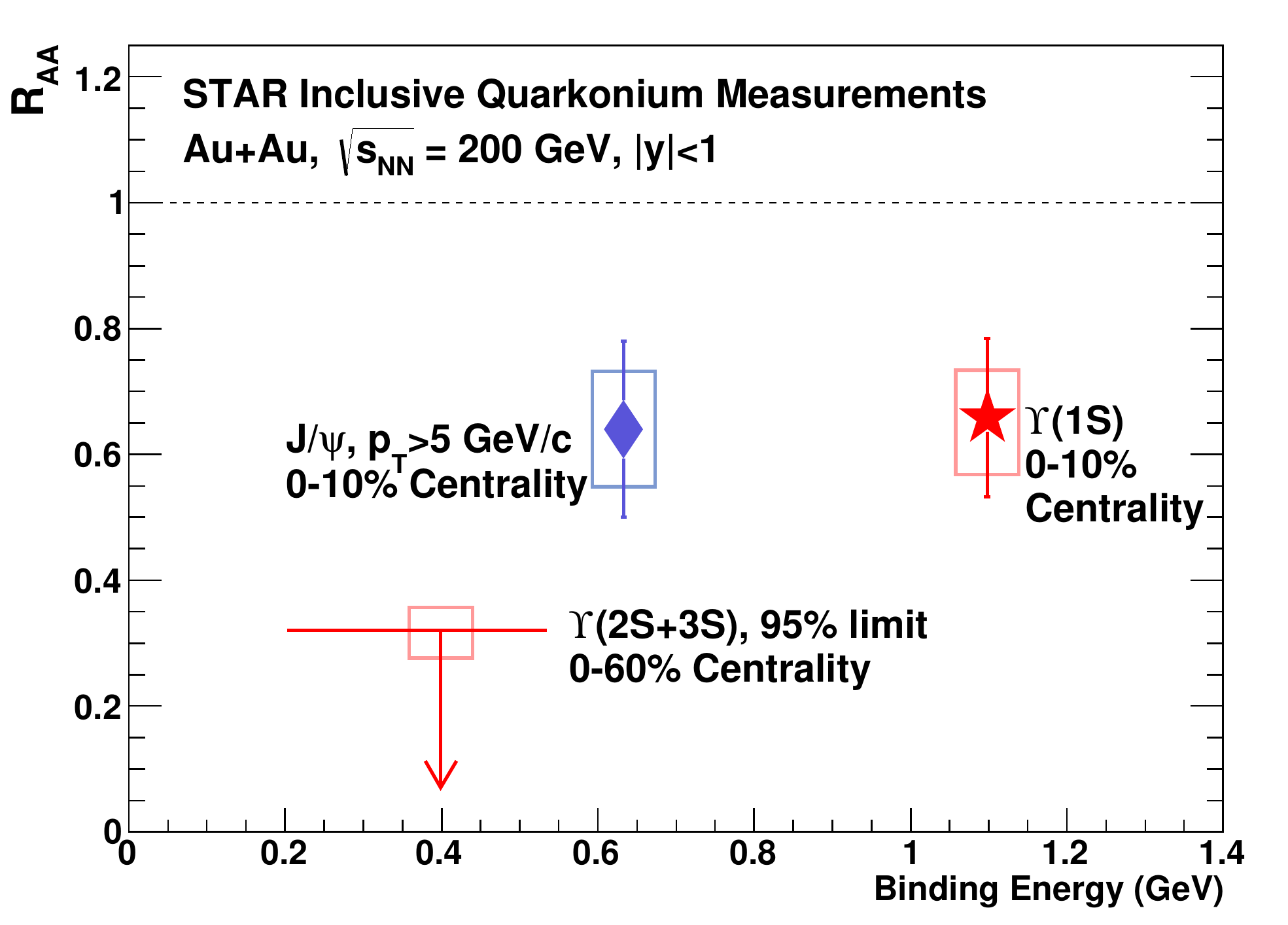}
		\includegraphics[width=0.49\linewidth]{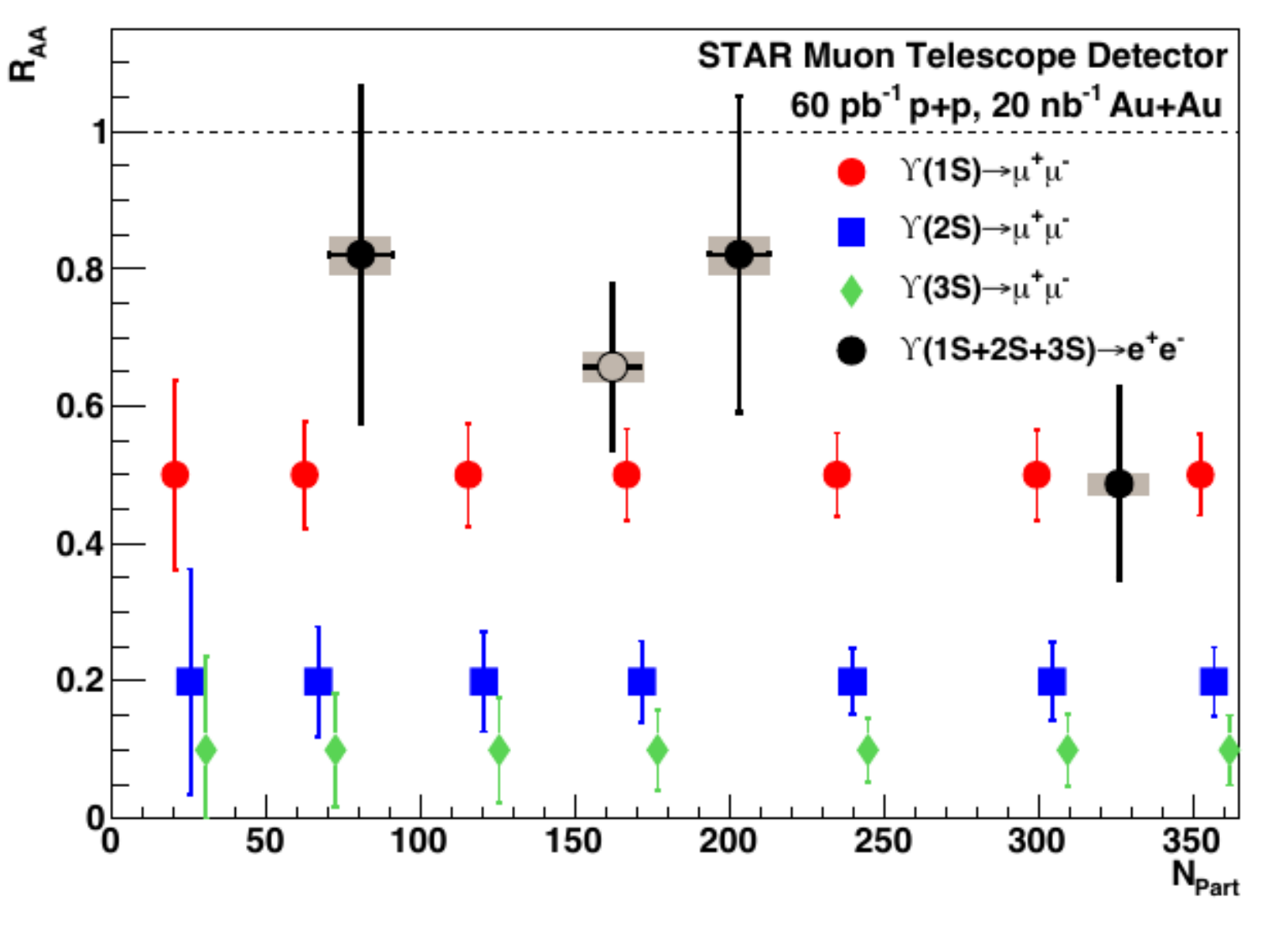}
		\caption{Left: STAR measurements of nuclear modification factors for different quarkonium states as a function of the binding energy (\cite{Adamczyk:2013poh}). Right: Projection of statistical uncertainties of $\Upsilon$ $R_{AA}$ measurement for different $\Upsilon$ states with the MTD detector. The figure also shows in black measured STAR $\Upsilon$ $R_{AA}$ \cite{Adamczyk:2013poh}.}
		\label{fig:bindingEn_UpsilonRaa_MTD}
\end{figure}

$R_{AA}$ as a function of quarkonium binding energies, for $\Upsilon(2S+3S)$ (0-60\%), high-$p_{T}$ J/$\psi$ (0-10\%) and $\Upsilon(1S)$ (0-10\%), are shown in the left panel of Fig.~\ref{fig:bindingEn_UpsilonRaa_MTD}. Suppressions of high-$p_{T}$ J/$\psi$ and $\Upsilon(1S)$ in the most central Au+Au collisions are at the same level, and the $\Upsilon(2S+3S)$ measurement suggests melting of these $\Upsilon$ state in Au+Au collisions at $\sqrt{s_{NN}} =$ 200 GeV.

\section{Outlook}
\label{sec:outlook}

STAR quarkonium measurements can be improved with two new detectors fully installed and taking data since 2014: Heavy Flavor Tracker (HFT) and Muon Telescope Detector (MTD). The HFT is a silicon vertex detector that will allow us to study the non-prompt J/$\psi$ production from B-meson decay, via topological reconstruction of displaced vertices.
The MTD enables muon identification in STAR and thus quarkonium measurements in the di-muon decay channel which is cleaner than the di-electron decay channel. Compared to electrons, muons do not originate from $\gamma$ conversion in the detector material and receive much less contribution from  Dalitz decays. Muons are also less affected by radiative loses in the detector material, therefore the di-muon channel has better mass resolution that is particularly important for separation of different $\Upsilon$ states. The right panel of Fig.~\ref{fig:bindingEn_UpsilonRaa_MTD} shows a projection for $R_{AA}$ of different $\Upsilon$ states measured with the MTD.

\section{Summary}
\label{sec:summary}

In this contribution we present STAR results on the J/$\psi$ and $\Upsilon$ production at mid-rapidity.
J/$\psi$ is measured in Au+Au collisions at different colliding energies, ranging from $\sqrt{s_{NN}} =$ 39 up to 200 GeV and in U+U collisions at $\sqrt{s_{NN}} =$ 193 GeV. Significant J/$\psi$ suppression is observed in central collisions, with no strong energy or colliding system dependence.
$\Upsilon$ results are shown in $p+p$, $d+Au$ and Au+Au collisions at $\sqrt{s_{NN}} =$ 200 GeV and in U+U collisions at $\sqrt{s_{NN}} =$ 193 GeV.
The strong suppression of high-$p_{T}$ J/$\psi$ and $\Upsilon$ in the most central 200 GeV Au+Au collisions, together with the indication of the complete suppression of $\Upsilon(2S)$ and $\Upsilon(3S)$, point to a presence of the Quark-Gluon Plasma.
With the new upgrades, STAR will continue its quarkonium physics program and will be able to perform more precise quarkonium measurements in next years.

\section*{Acknowledgements}

This publication was supported by the European social fund within the framework of realizing the project ,,Support of inter-sectoral mobility and quality enhancement of research teams at Czech Technical University in Prague'', CZ.1.07/2.3.00/30.0034 and by Grant Agency of the Czech Republic, grant No.13-20841S. 

\bibliographystyle{JHEP} 
\bibliography{biblio.bib}

\providecommand{\href}[2]{#2}\begingroup\raggedright\begin{thebibliography}{10}

\bibitem{Matsui:1986dk}
T.~Matsui and H.~Satz, {\it {$J/\psi$ Suppression by Quark-Gluon Plasma
  Formation}},  {\em Phys.Lett.} {\bf B178} (1986) 416.

\bibitem{Mocsy:2008eg}
A.~Mocsy, {\it {Potential Models for Quarkonia}},  {\em Eur.Phys.J.} {\bf C61}
  (2009) 705--710, [\href{http://arxiv.org/abs/0811.0337}{{\tt
  arXiv:0811.0337}}].

\bibitem{Mocsy:2007jz}
A.~Mocsy and P.~Petreczky, {\it {Color screening melts quarkonium}},  {\em
  Phys.Rev.Lett.} {\bf 99} (2007) 211602,
  [\href{http://arxiv.org/abs/0706.2183}{{\tt arXiv:0706.2183}}].

\bibitem{Rapp:2008tf}
R.~Rapp, D.~Blaschke, and P.~Crochet, {\it {Charmonium and bottomonium
  production in heavy-ion collisions}},  {\em Prog.Part.Nucl.Phys.} {\bf 65}
  (2010) 209--266, [\href{http://arxiv.org/abs/0807.2470}{{\tt
  arXiv:0807.2470}}].

\bibitem{Adamczyk:2013poh}
{\bf STAR} Collaboration, L.~Adamczyk et~al., {\it {Suppression of $\Upsilon$
  production in d+Au and Au+Au collisions at $\sqrt{s_{NN}}$=200 GeV}},  {\em
  Phys.Lett.} {\bf B735} (2014) 127--137,
  [\href{http://arxiv.org/abs/1312.3675}{{\tt arXiv:1312.3675}}].

\bibitem{Zhao:2010nk}
X.~Zhao and R.~Rapp, {\it {Charmonium in Medium: From Correlators to
  Experiment}},  {\em Phys.Rev.} {\bf C82} (2010) 064905,
  [\href{http://arxiv.org/abs/1008.5328}{{\tt arXiv:1008.5328}}].

\bibitem{Adamczyk:2012pw}
{\bf STAR} Collaboration, L.~Adamczyk et~al., {\it {Measurement of $J/\psi$
  Azimuthal Anisotropy in Au+Au Collisions at $\sqrt{s_{NN}}$ = 200 GeV}},
  {\em Phys.Rev.Lett.} {\bf 111} (2013), no.~5 052301,
  [\href{http://arxiv.org/abs/1212.3304}{{\tt arXiv:1212.3304}}].

\bibitem{Affolder:1999wm}
{\bf CDF} Collaboration, T.~Affolder et~al., {\it {Production of $\Upsilon(1S)$
  mesons from $\chi_b$ decays in $p\bar{p}$ collisions at $\sqrt{s} = 1.8$
  TeV}},  {\em Phys.Rev.Lett.} {\bf 84} (2000) 2094--2099,
  [\href{http://arxiv.org/abs/hep-ex/9910025}{{\tt hep-ex/9910025}}].

\bibitem{Ackermann:2002ad}
{\bf STAR} Collaboration, K.~Ackermann et~al., {\it {STAR detector overview}},
  {\em Nucl.Instrum.Meth.} {\bf A499} (2003) 624--632.

\bibitem{Anderson:2003ur}
M.~Anderson, J.~Berkovitz, W.~Betts, R.~Bossingham, F.~Bieser, et~al., {\it
  {The Star time projection chamber: A Unique tool for studying high
  multiplicity events at RHIC}},  {\em Nucl.Instrum.Meth.} {\bf A499} (2003)
  659--678, [\href{http://arxiv.org/abs/nucl-ex/0301015}{{\tt
  nucl-ex/0301015}}].

\bibitem{Llope:2012zz}
{\bf STAR} Collaboration, W.~Llope, {\it {Multigap RPCs in the STAR experiment
  at RHIC}},  {\em Nucl.Instrum.Meth.} {\bf A661} (2012) S110--S113.

\bibitem{Beddo:2002zx}
{\bf STAR} Collaboration, M.~Beddo et~al., {\it {The STAR barrel
  electromagnetic calorimeter}},  {\em Nucl.Instrum.Meth.} {\bf A499} (2003)
  725--739.

\bibitem{Llope:2003ti}
W.~Llope, F.~Geurts, J.~Mitchell, Z.~Liu, N.~Adams, et~al., {\it {The TOFp /
  pVPD time-of-flight system for STAR}},  {\em Nucl.Instrum.Meth.} {\bf A522}
  (2004) 252--273, [\href{http://arxiv.org/abs/nucl-ex/0308022}{{\tt
  nucl-ex/0308022}}].

\bibitem{Adamczyk:2013tvk}
{\bf STAR} Collaboration, L.~Adamczyk et~al., {\it {$J/\psi$ production at low
  $p_{T}$ in Au+Au and Cu+Cu collisions at $\sqrt{s_{NN}}=200$ GeV with the
  STAR detector}},  {\em Phys.Rev.} {\bf C90} (2014), no.~2 024906,
  [\href{http://arxiv.org/abs/1310.3563}{{\tt arXiv:1310.3563}}].

\bibitem{Adamczyk:2012ey}
{\bf STAR} Collaboration, L.~Adamczyk et~al., {\it {$J/\psi$ production at high
  transverse momenta in $p+p$ and Au+Au collisions at $\sqrt{s_{NN}} = 200$
  GeV}},  {\em Phys.Lett.} {\bf B722} (2013) 55--62,
  [\href{http://arxiv.org/abs/1208.2736}{{\tt arXiv:1208.2736}}].

\bibitem{Liu:2009nb}
Y.-p. Liu, Z.~Qu, N.~Xu, and P.-f. Zhuang, {\it {J/$\psi$ Transverse Momentum
  Distribution in High Energy Nuclear Collisions at RHIC}},  {\em Phys.Lett.}
  {\bf B678} (2009) 72--76, [\href{http://arxiv.org/abs/0901.2757}{{\tt
  arXiv:0901.2757}}].

\bibitem{Nelson:2012bc}
R.~Nelson, R.~Vogt, and A.~Frawley, {\it {Narrowing the uncertainty on the
  total charm cross section and its effect on the J/$\psi$ cross section}},
  {\em Phys.Rev.} {\bf C87} (2013), no.~1 014908,
  [\href{http://arxiv.org/abs/1210.4610}{{\tt arXiv:1210.4610}}].

\bibitem{Kikola:2011zz}
D.~Kikola, G.~Odyniec, and R.~Vogt, {\it {Prospects for quarkonia production
  studies in U + U collisions}},  {\em Phys.Rev.} {\bf C84} (2011) 054907,
  [\href{http://arxiv.org/abs/1111.4693}{{\tt arXiv:1111.4693}}].

\bibitem{Frawley:2008kk}
A.~D. Frawley, T.~Ullrich, and R.~Vogt, {\it {Heavy flavor in heavy-ion
  collisions at RHIC and RHIC II}},  {\em Phys.Rept.} {\bf 462} (2008)
  125--175, [\href{http://arxiv.org/abs/0806.1013}{{\tt arXiv:0806.1013}}].

\bibitem{Adare:2012bv}
{\bf PHENIX} Collaboration, A.~Adare et~al., {\it {$\Upsilon(1S+2S+3S)$
  production in $d+$Au and $p+p$ collisions at $\sqrt{s_{NN}}=200$ GeV and
  cold-nuclear matter effects}},  {\em Phys.Rev.} {\bf C87} (2013) 044909,
  [\href{http://arxiv.org/abs/1211.4017}{{\tt arXiv:1211.4017}}].

\bibitem{Arleo:2012rs}
F.~Arleo and S.~Peigne, {\it {Heavy-quarkonium suppression in p-A collisions
  from parton energy loss in cold QCD matter}},  {\em JHEP} {\bf 1303} (2013)
  122, [\href{http://arxiv.org/abs/1212.0434}{{\tt arXiv:1212.0434}}].

\bibitem{Strickland:2011aa}
M.~Strickland and D.~Bazow, {\it {Thermal Bottomonium Suppression at RHIC and
  LHC}},  {\em Nucl.Phys.} {\bf A879} (2012) 25--58,
  [\href{http://arxiv.org/abs/1112.2761}{{\tt arXiv:1112.2761}}].

\bibitem{Emerick:2011xu}
A.~Emerick, X.~Zhao, and R.~Rapp, {\it {Bottomonia in the Quark-Gluon Plasma
  and their Production at RHIC and LHC}},  {\em Eur.Phys.J.} {\bf A48} (2012)
  72, [\href{http://arxiv.org/abs/1111.6537}{{\tt arXiv:1111.6537}}].

\bibitem{Liu:2010ej}
Y.~Liu, B.~Chen, N.~Xu, and P.~Zhuang, {\it {$\Upsilon$ Production as a Probe
  for Early State Dynamics in High Energy Nuclear Collisions at RHIC}},  {\em
  Phys.Lett.} {\bf B697} (2011) 32--36,
  [\href{http://arxiv.org/abs/1009.2585}{{\tt arXiv:1009.2585}}].

\end{thebibliography}\endgroup

\end{document}